\documentclass[%
 reprint,
 amsmath,amssymb,
 aps,
pra,
]{revtex4-2}
\usepackage{graphicx}
\usepackage{dcolumn}
\usepackage{bm}

\usepackage[utf8]{inputenc}
\usepackage[T1]{fontenc}
\usepackage{booktabs, array, mathptmx, float, tabularx, booktabs, lipsum, amsmath,multirow}
\DeclareMathAlphabet{\mathcal}{OMS}{cmsy}{m}{n}
\DeclareSymbolFont{largesymbols}{OMX}{cmex}{m}{n}
\usepackage{siunitx, xcolor}
\usepackage[version=4]{mhchem}
\usepackage {nopageno} 
\graphicspath{{figs/}{figsgaoerb/}} 
\usepackage[colorlinks,linkcolor=blue,anchorcolor=blue,citecolor=blue]{hyperref}
\begin{document}
\renewcommand\arraystretch{1}
\preprint{APS/123-QED}
\title{Non-Hermitian dynamics of contextuality in $\mathcal{PT}$- and $\mathcal{APT}$-symmetry systems}

\author{Xuan Fan}
\author{Ya Xiao}%
 \email{xiaoya@ouc.edu.cn}
\author{Yongjian Gu}
 \email{yjgu@ouc.edu.cn}
\affiliation{%
 College of Physics and Optoelectronic Engineering, Ocean University of China, Qingdao City, Shandong Province, People's Republic of China, 266100\\
}%
\date{\today}

\begin{abstract}
In the past decades, researches on parity-time ($\mathcal{PT}$) and anti-parity-time ($\mathcal{APT}$) systems have garnered unprecedented attention, showcasing their various intriguing characteristics and promising potentiality in extending canonical Hermitian quantum mechanics. 
However, despite significant endeavors devoted to this new field of physics, non-Hermitian dynamics of contextuality still remains an uncharted region, either in $\mathcal{PT}$-symmetry or $\mathcal{APT}$-symmetry systems. 
Since contextuality has also been proven to be the core resource for quantum state discrimination (QSD) tasks, here we systematically investigate the novel performance of contextuality through QSD in both systems, taking mirror-symmetric three-state minimum error discrimination (MED) and maximum confidence discrimination (MCD) scenarios as two examples. The time evolution of contextuality in two scenarios and eight regimes (four regimes for each scenario) are comprehensively compared and analyzed, with the difference of initial states also considered. In the symmetry-unbroken regimes, our simulation shows periodic oscillations of contextuality for both MED and MCD scenarios, the period of which is state-independent but related to non-Hermiticity of the system. 
Both MED and MCD shows non-trivial recovery of contextuality exceeding its initial value in $\mathcal{PT}$ system, which is only existent for MCD in $\mathcal{APT}$ system. In the symmetry-broken regimes, the success probabilities of both scenarios start from a prompt decay at first, ending up with a stable value which is constantly $\frac{1}{3}$. Non-triviality is found only for MCD scenario in $\mathcal{PT}$ system, where the recovered contextuality exceeds its initial value. This work blazes the trail for studying non-Hermitian dynamics of contextuality, and also provides a new insight for the research of QSD performance in $\mathcal{PT}$ and $\mathcal{APT}$ systems, both of which play crucial role in quantum computation, communication, and cryptography.


\end{abstract}

\maketitle

\section{\label{Intro}Introduction}
Ever since its proposal at the end of last century\cite{bender1998real}, $\mathcal{PT}$-symmetry non-Hermitian system has been the focus of extensive interest for its non-trivial physical nature and unexpected potentiality in finding solutions for unsolvable problems in the conventional Hermitian quantum mechanics\cite{konotop2016nonlinear,el2018non}. Here, $\mathcal{P}=\left(\begin{array}{cc}
0 &  1 \\
1 &  0
\end{array}\right)$ is the parity reflection operator while the time-reversal operator $\mathcal{T}$ stands for complex conjugation. Despite its non-Hermiticity, Hamiltonians that satisfy $\mathcal{PT}$-symmetry $[\hat{H}_{\mathcal{PT}},\mathcal{PT}]=0$ can also ensure the reality of eigenvalue spectra in its symmetry-unbroken (PTS, $\mathcal{PT}$-symmetric) regime, just like those Hermitian Hamiltonians, which would require $\hat{H} = \hat{H}^{\dagger}$. However, the system will enter into the symmetry-broken (PTB, $\mathcal{PT}$-broken)  regime beyond the exceptional point (EP).
$\mathcal{APT}$-symmetry, as a counterpart for $\mathcal{PT}$-symmetry, has also caught researchers' attention considerably in the past years. Unlike $\mathcal{PT}$ systems, an $\mathcal{APT}$ system requires its Hamiltonian to be anti-commutative with the joint $\mathcal{PT}$ operator $\{\hat{H}_{\mathcal{APT}},\mathcal{PT}\}=0$. It is found that EP also exists in the $\mathcal{APT}$ system, separating its symmetry-unbroken (APTS, $\mathcal{APT}$-symmetric) and symmetry-broken (APTB, $\mathcal{APT}$-broken) regimes\cite{choi2018observation}. Despite that $\hat{H}_{\mathcal{APT}}=i\hat{H}_{\mathcal{PT}}$ holds one-to-one correspondent relation differing only by an imaginary number, their dynamics could still be of great difference. The unique feature of $\mathcal{APT}$ system has aroused many counter-intuitive phenomena, such as information flow\cite{kawabata2017information,chakraborty2019information,haseli2014non,wen2020observation}, coherence flow\cite{nair2021enhanced,fang2021experimental} and entanglement dynamics\cite{chen2014increase,fang2022entanglement,cen2022anti}. Both $\mathcal{PT}$ and $\mathcal{APT}$ experiments have been carried out in various physical quantum systems, including but not limited to 
optical\cite{ge2013antisymmetric,yang2017anti,konotop2018odd,wang2020experimental},
ion trap\cite{wang2021observation,ding2021experimental,ding2022information,bian2023quantum}, superconducting\cite{naghiloo2019quantum,partanen2019exceptional,agarwal2021pt}, and waveguide systems\cite{schnabel2017pt,qi2021encircling,qin2021quantum,liu2024floquet}.\par
Recently, much attention has been paid to the non-Hermitian version of quantum state discrimination (QSD), a central issue in Hermitian quantum mechanics, fundamental to many applications such as quantum metrology\cite{giovannetti2011advances,kurdzialek2023using}, key distribution\cite{bennett1992quantum,scarani2009security} and random number generation\cite{brask2017megahertz,i2022quantum}. 
In 2013, Benders et al. suggested the non-Hermitian evolution of two non-orthogonal input states into orthogonal states\cite{bender2013pt}, which is found faster than its Hermitian counterpart\cite{zheng2013observation}, indicating the possibility of finding a universal optimal discrimination for arbitrary non-orthogonal states\cite{chen2022quantum,wang2024demonstration}. 
However, previous works mainly focus on the state perspective, the non-Hermitian evolution of which leads to those non-trivial phenomena observed; while from a new, resource perspective, we would have a comprehensive understanding from a state-independent higher dimension, and thus might draw a more general conclusion across different QSD scenarios. 
Here we choose to use contextuality, a fundamental notion of quantum mechanics which serves as the key resource for various QSD tasks\cite{schmid2018contextual,zhang2022experimental,mukherjee2022discriminating,rossi2023contextuality}.
Our new perspective could also spark future researches to notice the non-Hermitian behavior of other quantum resources in their correspondent tasks.

In this paper, we theoretically investigate the non-Hermitian dynamical evolution of contextulaity in QSD tasks, where contextuality plays a key role. Various novel phenomena are found through our theoretical simulations. As for the symmetry unbroken regime, on the one hand, periodic oscillations of contextuality have been observed in both $\mathcal{PT}$- and $\mathcal{APT}$-symmetry systems for either MED or MCD scenarios. On the other hand, we find the phenomenon of stable value (PSV) in the symmetry broken regime, which is neither state-dependent nor non-Hermicity-relevant, approaching a universal constant $\frac{1}{3}$ in all cases. Noticeably, non-trivial recovery of contextuality is witnessed both in PTS regime for MED scenario, and in PTS, PTB, APTS regimes for MCD scenario. 
Our findings reveal novel phenomena in the time evolution of contextuality in $\mathcal{PT}$- and $\mathcal{APT}$  systems, and demonstrate the non-trivial contextual advantage in MED and MCD tasks even for non-Hermitian quantum systems. This work not only opens avenues for future study he dynamics of quantum contextuality in non-Hermitian systems, but also provides guides for further application of contextuality as an essential resource in various QSD scenarios as well as many other tasks that need them. 

\section{\label{Theory}THEORY}
The $2\times2$ $\mathcal{PT}$-symmetry non-Hermitian Hamiltonian for a single qubit generally takes the following form: 
\begin{equation}
    \begin{aligned}
        \hat{H}_{\mathcal{PT}} = \left(\begin{array}{cc}
r e^{i\varphi} &  s \\
s &  r e^{-i\varphi}
\end{array}\right) = r \cos\varphi\hspace{0.01cm} \mathbb{I}+s\hat{\sigma}(1,0,ia),
    \end{aligned}\label{HPT}
\end{equation}
where $a=\frac{r\sin\varphi}{s}$ quantifies non-Hermiticity of the $\mathcal{PT}$ system, and all the parameters $r$, $\varphi$ and $s$ here are real numbers controlling the non-Hermicity, with $\mathbb{I}$ denoting 2 × 2 identity matrix and $\hat{\sigma}[i]$ (here $i\in\{1,2,3\}$) the Pauli matrices:
\begin{equation}
    \begin{aligned}
        \hat{\sigma}[1] = \left(\begin{array}{cc}
0 &  1\\
1 &  0
\end{array}\right), \hat{\sigma}[2] = \left(\begin{array}{cc}
0 &  -i\\
i &  0
\end{array}\right), \hat{\sigma}[3] = \left(\begin{array}{cc}
1 &  0\\
0 &  -1
\end{array}\right).
    \end{aligned}
\end{equation}

\par
The eigenvalues for Hamiltonian $\hat{H}_{\mathcal{PT}}$ are given by 
\begin{equation}
    \begin{aligned}
       E_{\pm} = r\cos\varphi\pm\sqrt{s^2-r^2 \sin^2\varphi}=r\cos\varphi\pm s\sqrt{1-a^2}.
    \end{aligned}
\end{equation}\par
Here we define energy difference of the eigenvalues as
\begin{equation}
    \begin{aligned}
       2\omega=E_+ - E_- = 2s\sqrt{1-a^2}.
    \end{aligned}\label{omega}
\end{equation}\par
As can be seen, the value of $\omega$ is not affected by $r\cos\varphi$. Thus without loss of generality, we set $\varphi=\frac{\pi}{2}$ and $s=1$ in the following discussion, indicating a zero-valued real part for the diagonal terms of $ \hat{H}_{\mathcal{PT}}$, and a fixed value for its antidiagonal terms, in order to make the physical model more intuitive. Substitute those values into Eq. (\ref{HPT}) and (\ref{omega}) , we have
\begin{equation}
    \begin{aligned}
        \hat{H}_{\mathcal{PT}} = \left(\begin{array}{cc}
i r  &  1 \\
1 &  -i r
\end{array}\right) = \hat{\sigma}(1,0,ia),\hspace{0.2cm} 2\omega=2\sqrt{1-a^2}.
    \end{aligned}
\end{equation}\par
Given $a_1>1$, the two eigenvalues form a complex conjugate pair in the $\mathcal{PT}$ symmetry-broken regime. On the other hand, they become real numbers when where  $a<1$, which corresponds to the $\mathcal{PT}$ symmetry-unbroken regime. 
Thus we can deduce the time-evolution operator governed by $\hat{H}_{\mathcal{PT}}$:
\begin{equation}
    \begin{aligned}
       \hat{U}_{\scriptstyle{\mathcal{PT}}}&= \mathrm{\exp}(-i\hspace{0.05cm}\frac{\hat{H}_{\scriptscriptstyle\mathcal{PT}}}{\hbar}\hspace{0.05cm}t)\\
        &=\frac{1}{\sin\alpha}\left[\begin{array}{cc}\scriptstyle{
\sin(\alpha+\sqrt{1-a^2}\hspace{0.04cm}t)} &  \scriptstyle{i \sin(\sqrt{1-a^2}\hspace{0.04cm}t)}\\
\scriptstyle{i \sin(\sqrt{1-a^2}\hspace{0.04cm}t)} &  \scriptstyle{\sin(\alpha-\sqrt{1-a^2}\hspace{0.04cm}t)}
\end{array}\right],
    \end{aligned}
\end{equation}
where we have chosen to work in units where the reduced Planck’s constant $\hbar=1$ and assumed $\alpha=\arccos{a}$.

Similarly, we can further simplify the generalized form of a single-qubit APT-symmetry Hamiltonian as follows:
\begin{equation}
    \begin{aligned}
        \hat{H}_\mathcal{APT} = \left(\begin{array}{cc}
r e^{i\varphi} &  i s \\
i s &  -r e^{-i\varphi}
\end{array}\right)\xrightarrow{\varphi=0,\hspace{0.1cm}s=1}
\left(\begin{array}{cc}
b &  i \\
i &  -b
\end{array}\right),
    \end{aligned}
\end{equation}
where $b=\frac{r\sin\varphi}{s}$ quantifies non-Hermiticity of $\mathcal{APT}$ system. The time-evolution operator for $\hat{H}_{\mathcal{APT}}$ can be written as
\begin{equation}
    \begin{aligned}
       \hat{U}_{\scriptstyle{\mathcal{APT}}}&= \mathrm{\exp}(-i\hspace{0.05cm}\frac{\hat{H}_{\scriptscriptstyle\mathcal{APT}}}{\hbar}\hspace{0.05cm}t)=\left[\begin{array}{cc}A-i B &  C\\
C &  A+i B
\end{array}\right],
    \end{aligned}
\end{equation}
where $A, B, C$ are respectively given by\par
(1) for $b>1,$
\begin{equation}
    \begin{aligned}
       A &= \cos({\sqrt{b^2-1}\hspace{0.04cm}\textstyle t}), B = \frac{b}{\sqrt{b^2-1}}\sin{(\sqrt{b^2-1}\hspace{0.04cm}t)},\\
       C &= \frac{1}{\sqrt{b^2-1}}\sin({\sqrt{b^2-1}\hspace{0.04cm}t}).
    \end{aligned}
\end{equation}\par
(2) for $b<1,$
\begin{equation}
    \begin{aligned}
       A &= \cosh({\sqrt{1-b^2}\hspace{0.04cm}\textstyle t}), B = \frac{b}{\sqrt{1-b^2}}\sinh{(\sqrt{1-b^2}\hspace{0.04cm}t)},\\
       C &= \frac{1}{\sqrt{1-b^2}}\sinh({\sqrt{1-b^2}\hspace{0.04cm}t}).
    \end{aligned}
\end{equation}\par
(3) for $b=1,$
\begin{equation}
    \begin{aligned}
       A = 1, B = t,
       C &= t.
    \end{aligned}
\end{equation}
\section{\label{Simulation}RESULTS OF SIMULATIONS}
State discrimination is an essential task in quantum cryptography, communication and computation, and has been broadly studied for its wide application. Due to the impossibility of perfectly discriminating non-orthogonal states, many strategies have been put forward, among which MED and MCD have been taken as two important figure-of-merits. 
\par
In this section we will briefly introduce MED and MCD discrimination of three mirror-symmetric states shown below:
\begin{equation}
    \begin{aligned}
&|\mathcal{\psi}_1\rangle=\cos\theta|0\rangle+\sin\theta|1\rangle,\\
        &|\psi_2\rangle=\cos\theta|0\rangle-\sin\theta|1\rangle,\\
        &|\psi_3\rangle=|0\rangle,
    \end{aligned}\label{State}
\end{equation}
which are taken with prior probabilities $p_1=p_2=p$ and $p_3=1-2p$, where $0<p<1/2$ and $0<\theta<\pi/2$. \par
\subsection{\label{MEDPTAPT}Non-Hermitian Evolution of Contextuality in MED}
The MED strategy aims at finding a set of measurements which minimizes the probability of making a wrong guess in figuring out which state is given. For mirror-symmetric three-state MED, its success probability is given by $\sum_0^ip_i\rm Tr( \rho_\textit{i} \it E_i\rm )$, where $\rho_i = |\psi_i\rangle\langle\psi_i|$ is the density matrix for each state, and $E_i$ stands for the optimal measurement corresponding to that state, the best strategy of which is given in \cite{andersson2002minimum}.
From quantum theory we can easily obtain its contextual bound
\vspace{-0.2cm}
\begin{equation}
    \begin{aligned}
        S_Q &= p\rm Tr( \rho_1 \textit{E}_1)+p\rm Tr( \rho_2 \textit{E}_2)+(1-2p)\rm Tr( \rho_3 \textit{E}_3)\\
        &=  \begin{cases}
        p(1+\sin2\theta), &\hspace{-0.2cm}  \scriptstyle p\geq \frac{1}{2+\cos\theta(\cos\theta+\sin\theta)},\\
         \frac{(1-2p)(p\sin^2\theta+1-2p-p\cos^2\theta)}{2+\cos\theta(\cos\theta+\sin\theta)}, &\hspace{-0.2cm} \scriptstyle p< \frac{1}{2+\cos\theta(\cos\theta+\sin\theta)}.\\
    \end{cases}
    \end{aligned}
\end{equation}\par
For the classical counterpart, its upper bound can be deduced from the non-contextual hidden variable theory\cite{mukherjee2022discriminating}
\vspace{-0.1cm}
\begin{equation}
    \begin{aligned}
        &S_\lambda = \begin{cases}
        1-(1-2p)\cos^2\theta-p\cos^22\theta, & p\geq \frac{1}{3},\\
        1-p\cos^2\theta-p\cos^22\theta, & p< \frac{1}{3}.\\
    \end{cases}
    \end{aligned}
\end{equation}
\vspace{0.1cm}
\par
\begin{figure}[h]
\begin{center}
\includegraphics[width=1\columnwidth]{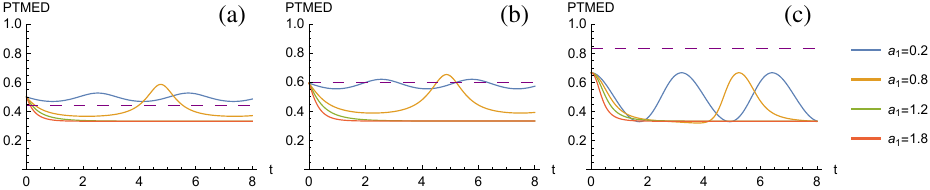}
\caption{Time evolution of success probability for MED strategy when the state evolves in the $\mathcal{PT}$-symmetry system. (a) $p=\frac{1}{3}, \theta=\frac{\pi}{12};$ (b) $p=\frac{1}{3}, \theta=\frac{\pi}{7};$ (c) $p=\frac{1}{3}, \theta=\frac{\pi}{3}$. In each subplot, curves of different colors represent $a_1$ of different values, while the dashed line in purple stands for the classical upper bound given current $\{p,\theta\}$.}
\label{PTMED}
\end{center}
\end{figure}

We will first discuss about the non-Hermitian evolution of contextuality in MED scenario. Fig. \ref{PTMED} demonstrates the evolution of contextuality in MED scenario over time $t$. Here Fig. \ref{PTMED}(a) represents the case of the following three initial states: $|\mathcal{\psi}_1\rangle=\frac{\sqrt{3}+1}{2\sqrt{2}}|0\rangle+\frac{\sqrt{3}-1}{2\sqrt{2}}|1\rangle,
        |\psi_2\rangle=\frac{\sqrt{3}+1}{2\sqrt{2}}|0\rangle-\frac{\sqrt{3}-1}{2\sqrt{2}}|1\rangle,
        |\psi_3\rangle=|0\rangle$, with prior probability of each state $p=\frac{1}{3}$, while Fig. \ref{PTMED}(b) and (c) stand for states with $\{p,\theta\}$ as $\{\frac{1}{3},\frac{\pi}{7}\}$ and $\{\frac{1}{3},\frac{\pi}{3}\}$.\par
As can be seen in Fig. \ref{PTMED}, the oscillation of contextuality occurs when $a_1<1$ (PTS regime), while the phenomenon of PSV occurs when $a_1>1$ (PTB regime). Generally in PTS regime, the oscillation period  increases together with $a_1$, with the peak value of recovery getting higher at the same time(see blue and yellow curves). Noticeably the recovery of contextuality sometimes exceeds even the initial value, especially in Fig. \ref{PTMED}(b), for example, where the oscillation in PTS regime induces a violation of the classical bound which is not reached by the initial value (see dashed purple line), showing the non-triviality of recovery. As for PTB regime, the final value for PSV is both state-independent and $a_1$-irrelevant, with the time needed to reach that value (always $\frac{1}{3}$) getting shortened as $a_1$ increases (see green and red curves). \par
\begin{figure}[t]
\begin{center}
\includegraphics[width=1\columnwidth]{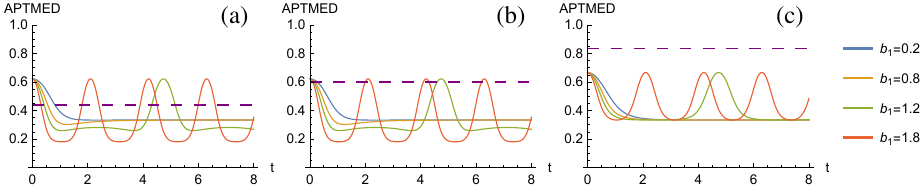}
\caption{Time evolution of success probability for MED strategy when the state evolves in the $\mathcal{APT}$-symmetry system. (a) $p=\frac{1}{3}, \theta=\frac{\pi}{12};$ (b) $p=\frac{1}{3}, \theta=\frac{\pi}{7};$ (c) $p=\frac{1}{3}, \theta=\frac{\pi}{3}$. In each subplot, curves of different colors represent $b_1$ of different values, while the dashed line in purple stands for the classical upper bound given current $\{p,\theta\}$.}
\label{APTMED}
\end{center}
\end{figure}

With the same set of initial states as shown in its $\mathcal{PT}$ counterpart, the dynamical characteristics of contextuality in $\mathcal{APT}$-symmetry system are depicted in Fig. \ref{APTMED},  with those subplots standing for $\{p,\theta\}$ as $\{\frac{1}{3},\frac{\pi}{12}\}$, $\{\frac{1}{3},\frac{\pi}{7}\}$ and $\{\frac{1}{3},\frac{\pi}{3}\}$. \par
Different from those $\mathcal{PT}$ cases, the oscillation of contextuality in $\mathcal{APT}$ regime occurs when $b_1>1$ (APTS regime), and the phenomenon of PSV occurs when $b_1<1$ (APTB regime), as Fig. \ref{APTMED} shows. 
Unlike what we see from those $\mathcal{PT}$ cases, the recovery of contextuality in APTS regime is a trivial process. The oscillation can at its best goes back to the initial value at $t=0$, which is $b_1$-irrelevant. Thus whether the recovered contextuality can violate the classical bound is predetermined by its initial value, a state-dependent variable. The minimum value, on the other hand, gets decreased as $b_1$ increases, showing a worse performance of MED in APTS regime, compared with either PTS or APTB regimes. Both the APTS period of oscillation, and the APTB time for decay decrease with an increasing $b_1$, with the final value of PSV also approaching $\frac{1}{3}$. 
\hspace{-2cm}
\subsection{\label{MCDPTAPT}Non-Hermitian Evolution of Contextuality in MCD}
MCD strategy, on the other hand, yields the maximal success probability (confidence) of correctly identifying one particular state out of a set of linearly dependent states. The highest confidence allowed by quantum and classical theories for mirror-symmetric states with $\{p,\theta\}$ are as follows\cite{croke2006maximum,mukherjee2022discriminating}:
\begin{equation}
    \begin{aligned}
        \mathrm{(Quantum)\hspace{0.2cm}}C_Q = \frac{p_1 Tr[\rho_1E_1]}{Tr[\rho E_1]}
        =  \frac{1+2p\cos\theta}{2-4p\sin^2\theta},\\
         \mathrm{(Classical)\hspace{0.2cm}}C_\lambda = \frac{1}{1+\cos^22\theta+(\frac{1}{p}-2)\cos^2\theta}.
    \end{aligned}
\end{equation}
\par
For the non-Hermitian dynamics of contextuality in MCD scenario, we will also start with its evolution in the $\mathcal{PT}$ system. Here the three subplots stand for initial states with $p=\frac{1}{3}$ and $\theta=\frac{\pi}{12}$(a),  $\theta=\frac{\pi}{7}$(b), and $\theta=\frac{\pi}{3}$(c), respectively.\par
\begin{figure}[h]
\begin{center}
\includegraphics[width=1\columnwidth]{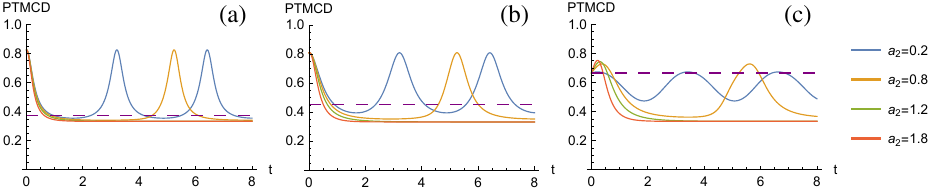}
\caption{Time evolution of success probability for MCD strategy when the state evolves in the $\mathcal{PT}$-symmetry system. (a) $p=\frac{1}{3}, \theta=\frac{\pi}{12};$ (b) $p=\frac{1}{3}, \theta=\frac{\pi}{7};$ (c) $p=\frac{1}{3}, \theta=\frac{\pi}{3}$. In each subplot, curves of different colors represent $a_2$ of different values, while the dashed line in purple stands for the classical upper bound given current $\{p,\theta\}$.}
\label{PTMCD}
\end{center}
\end{figure}
Fig. \ref{PTMCD} depicts the time evolution of MCD's maximal success probability (confidence) in the $\mathcal{PT}$ system. The PTS regime for MCD scenario, i.e. where the oscillation of contextuality occurs, is conditioned by $a_2<1$, and $a_2>1$ denotes the PTB regime, where PSV can be observed, same as its MED counterpart. The detailed patterns of both regimes are also similar, e.g. the dependence of their PTS oscillation period or PTB decay time on system's non-Hermiticity, and the constant stable value $\frac{1}{3}$ in PTB regime. However, MED only has a limited region of non-trivial recovery in PTS regime, while the MCD non-trivial recovery of contextuality covers even the broken regime. As can be clearly seen in Fig. \ref{PTMCD}(c), whatever regime it is, the higher $a_2$ gets, the better the recovery of contextuality is, and the faster the peak of recovery is reached. \par
\begin{figure}[h]
\begin{center}
\includegraphics[width=1\columnwidth]{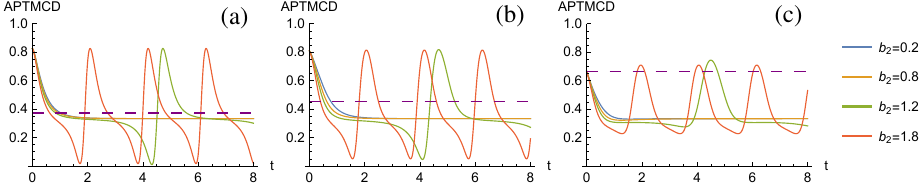}
\caption{Time evolution of success probability for MCD strategy when the state evolves in the $\mathcal{APT}$-symmetry system. (a) $p=\frac{1}{3}, \theta=\frac{\pi}{12};$ (b) $p=\frac{1}{3}, \theta=\frac{\pi}{7};$ (c) $p=\frac{1}{3}, \theta=\frac{\pi}{3}$. In each subplot, curves of different colors represent $b_2$ of different values, while the dashed line in purple stands for the classical upper bound given current $\{p,\theta\}$.}
\label{APTMCD}
\end{center}
\end{figure}
Last but not least, as for the performance of MCD strategy in the $\mathcal{APT}$ system, the overall pattern is also similar to its MED counterpart, with two remarkable differences worth discussing. First and most noticeable is its periodic oscillation with noteworthy unsymmetry to its extreme values in the APTS regime. Another interesting difference lies in the fact that for MCD strategy, as Fig. \ref{APTMCD}(c) demonstrates, the non-trivial recovery of contextuality also exists in APTS regime, which can only be found in PTS regime for MED. 

\section{OPTICAL IMPLEMENTATION}
\section{\label{summary}Conclusions}
In this paper, the purpose of our work is to elucidate the novel non-Hermitian performance of contextuality and its great potentiality in addressing fundamental issues in conventional Hermitian quantum mechanics, such as QSD. This work thus also serves as an extension of a series of studies on the intriguing features of QSD in non-Hermitian systems, which is fundamental for many quantum information tasks. Previous studies focus upon this issue mostly by trying different strategies, such as MED, and unambiguous discrimination(UD). Our work, on the other hand, takes a novel perspective into this issue by considering the time evolution of contextuality, a key quantum resource for QSD. We comprehensively studied the non-Hermitian performance of contextuality, the existence of which denotes advantage for QSD tasks, by studying the evolution of success probabilities  for two famous state discrimination scenarios: MED and MCD. Altogether eight cases are taken into consideration, with four regimes (PTS, PTB, APTS, APTB) simulated for each scenario. \par

We systematically simulated the non-Hermitian evolution of contextuality in both $\mathcal{PT}$ and $APT$ systems, taking its performance in QSD tasks as an example. By comparing different non-Hermitian regimes, QSD strategies, initial states and values of non-Hermiticity, the novel phenomena of contextual evolution are found and discussed in detail.
We observe the oscillation of contextuality in the symmetry-unbroken regimes for both $\mathcal{PT}$ and $\mathcal{APT}$ systems, and its decay in the symmetry-broken regimes. Both oscillation period and decay time has a high dependence on the non-Hermiticity of the system. In symmetry-broken regimes, a special phenomenon of PSV is discovered, where the final stable values for both scenarios fall onto a system-independent constant. Generally speaking, for a $\mathcal{PT}$-symmetry system,  the performance of contextuality in the PTS regime is always better than that in the PTB regime; While for an $\mathcal{APT}$-symmetry system, the minimum value in the APTS regime might be even lower than its APTB counterpart, denoting a possibly worse performance under symmetry-unbroken cases. In the PTS regime, nontrivial recovery of contextuality is witnessed for both MED and MCD strategies, where contextual oscillation might exceed its initial value. While for APTS and PTB cases, nontrivial recovery of contextuality exists only in the MCD scenario.\par
Our work unveils the non-trivial performance of contextuality in both MED and MCD scenarios, showing potential applicability of contextuality in non-Hermitian QSD tasks. The perspective used in this study is new to not only QSD but also many other tasks, and thus would encourage more research upon the non-Hermitian behavior of quantum resources in various applications. To our knowledge, this paper also denotes the first step into the theoretical exploration of contextuality in both $\mathcal{PT}$ and  $\mathcal{APT}$ systems, from which future ingenious studies on the non-Hermitian dynamics of other important quantum correlations might be inspired. 

\section{ACKNOWLEDGEMENT}
This work was supported by the National Natural Science Foundation Regional Innovation and Development Joint Fund (Grant No. U19A2075), the National Natural Scicence Foundation of China (Grant No. 12004358), the Fundamental Research Funds for the Central Universities (Grants No. 202041012 and No. 841912027), the Natural Science Foundation of Shandong Province of China (Grant No. ZR2021ZD19), and the Young Talents Project at Ocean University of China (Grant No. 861901013107).
\appendix
\bibliography{universalsample}

\end{document}